\documentclass[a4paper]{article}
\usepackage{ISCSLP2026}
\usepackage{ifthen}
\usepackage{times}
\usepackage[hidelinks]{hyperref}
\usepackage{graphicx}
\usepackage{amsmath,amssymb}
\usepackage{booktabs}
\usepackage{multirow}
\usepackage{array}
\newboolean{blind}
\setboolean{blind}{false}

\title{CF-AVSS: A Coarse-to-Fine Audio-Visual Speech Separation Framework with Target-Speaker Refinement and Post-Fusion for Real-World AVSE Challenge}

\name{
  \ifthenelse{\boolean{blind}}{Anonymous to ISCSLP}
  {Longjie Luo$^1$, Zhongshu Hou$^1$, Yuepeng Li$^1$, Gang Yang$^{1,2,*}$, Shidong Shang$^1$, Fan Zhou$^2$}
}

\address{
  \ifthenelse{\boolean{blind}}{Anonymous to ISCSLP}
  {$^1$Tencent Ethereal Audio Lab, Tencent Corporation, China \\
   $^2$University of Electronic Science and Technology of China, China}
}

\email{
  \ifthenelse{\boolean{blind}}{Anonymous to ISCSLP}
  {drakeluo@tencent.com, zhongshuhou@tencent.com,
   felixypli@tencent.com, gangy@std.uestc.edu.cn, simeonshang@tencent.com, 
   fan.zhou@uestc.edu.cn}
}

\begin{document}

\maketitle

\begin{abstract}
Audio-visual speech enhancement (AVSE) leverages visual cues to recover target speech from mixtures corrupted by reverberation, noise, and overlapping speakers. We propose a coarse-to-fine two-stage AVSE framework. In the first stage, a progressive SPAV-TFGridNet is extended to two synchronized visual streams to produce visually ordered coarse estimates by progressively suppressing interfering speech while preserving target reverberation and shared background noise. In the second stage, a TFGridNet-based target speaker refinement network takes the original mixture and both coarse estimates as input to further recover target speech and improve signal quality. The coarse and refined estimates are finally combined using development-set-calibrated fusion. On the ISCSLP 2026 Real-World AVSE Challenge, our system tied for first place on Track 1 and ranked second on Track 2 in OVRL, while achieving the lowest CER on both tracks at 12.31\% and 14.79\%, respectively.

\end{abstract}


\begingroup
\renewcommand{\thefootnote}{}
\footnotetext{\textsuperscript{*}Corresponding author. Work done during an internship at Tencent.}
\addtocounter{footnote}{-1}
\endgroup

\section{Introduction}

Audio-visual speech enhancement (AVSE) combines acoustic observations with
visual speech cues to recover desired speech from adverse recordings~\cite{qian2025sav,ren2024robust, chao2025leveraging,ren2025bav}. Lip
motion and facial articulation provide target-dependent information that does
not suffer from acoustic noise~\cite{son2017lip}, so AVSE can improve communication, automatic
speech recognition, and human-machine interaction when reverberation, noise,
or concurrent speakers make audio-only processing unreliable~\cite{ephrat2018looking}.

Most AVSE systems encode a noisy mixture and one target face track, align the
two feature streams, and estimate one enhanced waveform~\cite{gao2021visualvoice}. Time-domain
separators and time-frequency masking networks commonly follow this paradigm:
an acoustic encoder represents the mixture, a visual encoder represents mouth
motion, a fusion module injects the visual cue into the acoustic stream, and a
decoder reconstructs the target signal. This direct mapping works well for additive
simulations and clean frontal videos, but it leaves several problems in real
recordings. Natural overlap does not follow ideal additive assumptions; rooms
introduce long reverberant tails and device-dependent responses; and imperfect
faces, side views, occlusion, motion blur, low resolution, frozen frames, or
audio-visual desynchronization can weaken visual conditioning. Aggressive
one-step enhancement may also suppress weak target phonemes while removing
interference, which creates a conflict between perceptual quality,
intelligibility, and identity preservation.

The ISCSLP 2026 Real-World AVSE Challenge was designed to bridge the gap between clean-video benchmarks and practical deployment conditions~\cite{organizers2026iscslp2026realworldaudiovisual, yang2026identityfaithfulaudiovisualtargetspeaker}. Track~1 evaluates naturally captured multi-speaker recordings with spontaneous speech overlap, room reverberation, background noise, and device-dependent acoustic responses. Track~2 further degrades the visual modality through occlusion, low resolution, frozen or dropped frames, and audio-visual desynchronization, while also introducing far-field recordings captured at a distance of 3~m. Each track contains real \emph{mix} scenes without isolated acoustic references and reference-available \emph{remix} scenes constructed from separately recorded sources. These conditions require a system to maintain reliable target-speaker association when either the acoustic or visual evidence becomes unreliable. The evaluation consequently combines intrusive fidelity metrics, non-intrusive quality estimators, recognition accuracy, and speaker similarity, emphasizing balanced performance rather than optimization of a single enhancement criterion.

In this paper, we address these conditions with coarse separation followed by
fine extraction. We first extend SPAV-TFGridNet \cite{hou2024sir} from a
single visually conditioned output to two jointly estimated, visually ordered
outputs. Progressive supervision removes reverberant non-target speakers in
stages while preserving the target room response and common noise. We then
use a TFGridNet-based\cite{wang2023tf} refinement model to extract cleaner target
speech from the original mixture, guided by both coarse estimates. An asymmetric over-suppression loss discourages the removal of weak target components, while
PESQ-oriented and adversarial losses complement SI-SDR and compressed-spectral
reconstruction. Finally, we select one linear-fusion coefficient from the
development set and keep it fixed for test inference. The resulting system
obtained an official OVRL of 2.86 on both tracks, corresponding to joint first
place on Track~1 and second place on Track~2. Its CER of 12.31\% and 14.79\%
was the best submitted result on Track~1 and Track~2, respectively.

\begin{figure*}[t]
  \centering
  \includegraphics[width=0.94\textwidth]{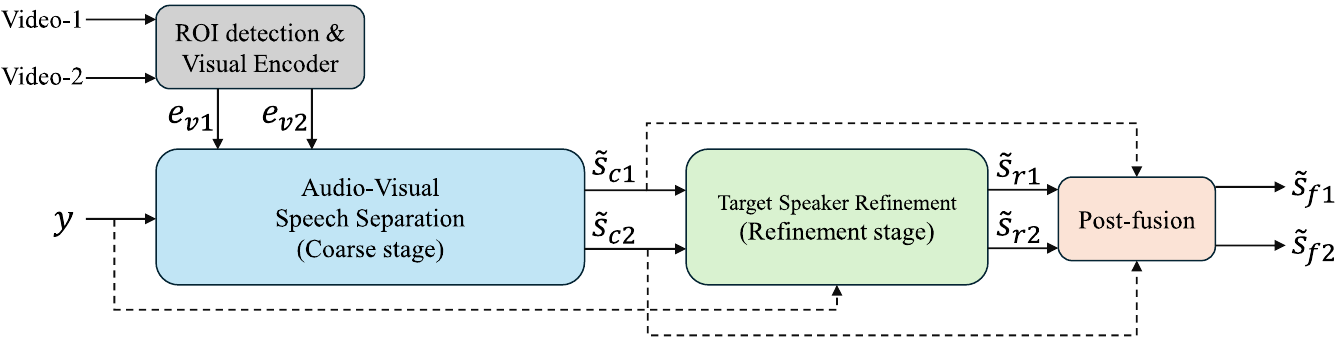}
  \caption{Overview of the proposed two-stage AVSE method.}
  \label{fig:overview}
\end{figure*}

\section{Preliminary}
\label{sec:preliminary}



\subsection{Problem formulation}

We consider a single-channel noisy and reverberant mixture containing $M\geq2$ speakers,
\begin{equation}
    y=\sum_{m=1}^{M}u_m+n,
    \qquad
    u_m=h_m*s_m,
    \label{eq:mixture}
\end{equation}
where $s_m$ is the clean speech of speaker $m$, $h_m$ is its room impulse response, $u_m$ is the corresponding reverberant speech, and $n$ denotes background noise. Given the mixture $y$ and two visual streams $(v_1,v_2)$, our system estimates two enhanced signals $(\tilde{s}_{f1},\tilde{s}_{f2})$, each aligned with its corresponding visible speaker, while suppressing the remaining interfering speakers.

\section{Method}
\label{sec:method}
In this section, we first describe the audio-visual representations used to associate the mixture with two target speakers. We then introduce the progressive coarse AVSS stage, which suppresses competing speakers while preserving target reverberation and shared background noise. Next, we present the contextual TSR network for recovering cleaner target signals from the original mixture under the guidance of the coarse estimates. Finally, we describe the development-set-calibrated post-fusion strategy that combines the complementary information retained by the coarse and refined outputs.

\subsection{System overview}

As shown in Fig.~\ref{fig:overview}, given an observed mixture $y$ and two synchronized video streams, the visual encoder extracts target-specific features $e_{v_i}$, where $i\in\{1,2\}$. Conditioned on these features, the coarse audio-visual speech separation (AVSS) stage suppresses competing speech and generates visually ordered estimates $\tilde{s}_{c_i}$, keeping each estimate aligned with its corresponding visual stream. The TFGridNet-based target speaker refinement (TSR) stage then jointly processes $y$ and both coarse estimates to produce refined estimates $\tilde{s}_{r_i}$, which are combined through development-set-calibrated post-fusion to obtain the final outputs $\tilde{s}_{f_i}$.

\subsection{Visual representation}

For each video stream, we apply the official AV-HuBERT preprocessing pipeline, which detects 68 facial landmarks, aligns the face to a canonical mean shape, and crops a mouth-centered region of interest (ROI) \cite{shi2022learning}. The ROI sequence is then processed by AV-HuBERT Base to directly extract visual speech features $e_{v_i}$ at 25 frames per second, where $i\in\{1,2\}$. The two visual features $e_{v_1}$ and $e_{v_2}$, together with the mixture $y$, are fed to the coarse separation stage while retaining their correspondence to the input video streams.

\begin{figure}[t]
  \centering
  \includegraphics[width=\columnwidth]{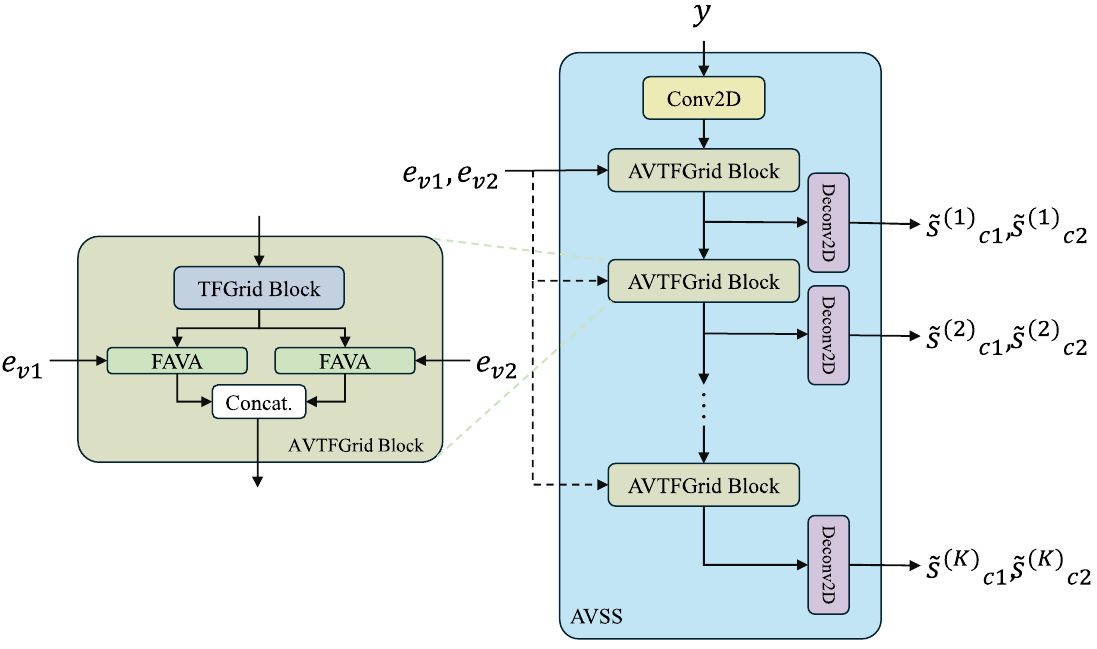}
  \caption{Architecture of the progressive coarse AVSS stage.}
  \label{fig:stage1}
\end{figure}

\subsection{Progressive coarse separation}

As shown in Fig.~\ref{fig:stage1}, the coarse AVSS stage suppresses competing speech while preserving target reverberation and shared background noise. We adapt SIR-Progressive Audio-Visual TF-GridNet \cite{hou2024sir} to jointly estimate two visually ordered targets. After Conv2D encoding, the mixture $y$ is processed by $K$ cascaded AVTFGrid blocks, each comprising a TFGrid block and two feature-wise audio-visual attention (FAVA) modules conditioned on $e_{v1}$ and $e_{v2}$~\cite{vaswani2017attention}; their internal structures follow \cite{hou2024sir}. The $k$-th decoder predicts
\begin{equation}
    \left(\tilde{s}_{c1}^{(k)},\tilde{s}_{c2}^{(k)}\right)
    =F_{\mathrm{sep}}^{(k)}(y,e_{v1},e_{v2}),
    \qquad k=1,\ldots,K,
    \label{eq:stage1}
\end{equation}
where $\tilde{s}_{ci}=\tilde{s}_{ci}^{(K)}$ is the final coarse estimate for visual stream $i$. Tying each output index to its visual stream avoids permutation ambiguity during both training and inference.

For target $i$, let $q_i=\sum_{m\neq i}u_m$ denote the competing reverberant speech. The final coarse target is
\begin{equation}
    c_i=u_i+n,
    \label{eq:coarse-target}
\end{equation}
which retains the complete reverberant target and shared background noise while excluding competing speakers. Since $n$ is shared by both targets, we impose no mixture-consistency constraint requiring $\tilde{s}_{c1}+\tilde{s}_{c2}=y$.

Following the progressive training strategy in \cite{hou2024sir}, the target at level $k$ is
\begin{equation}
    c_i^{(k)}=c_i+\gamma_k q_i,
    \qquad
    1\geq\gamma_1>\cdots>\gamma_K=0,
    \label{eq:progressive}
\end{equation}
where $\gamma_k$ follows the stage-wise SIR schedule. Thus, earlier levels retain more interfering speech, whereas the final level reaches the background-preserving coarse target.

For a complex STFT coefficient $Z$, we define power compression as
$C_p(Z)=|Z|^p\exp(j\angle Z)$. All progressive outputs are supervised by
\begin{align}
    \mathcal{L}_{\mathrm{sep}}
    =\sum_{k=1}^{K}w_k\sum_{i=1}^{2}\Big[
    &-\lambda_s\mathrm{SI\mbox{-}SDR}
    \left(\tilde{s}_{ci}^{(k)},c_i^{(k)}\right) \notag\\
    &+\lambda_c
    \left\|C_p\left(\tilde{S}_{ci}^{(k)}\right)
    -C_p\left(C_i^{(k)}\right)\right\|_2^2
    \Big],
    \label{eq:stage1loss}
\end{align}
where $\tilde{S}_{ci}^{(k)}$ and $C_i^{(k)}$ are the STFTs of the prediction and target, respectively, and $w_k$ is the level weight. The compressed spectral term preserves absolute scale and emphasizes low-energy components \cite{li2021power}.

\begin{figure}[t]
  \centering
  \includegraphics[width=0.4\columnwidth]{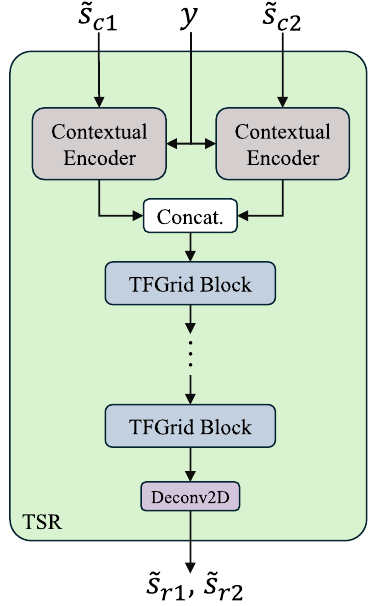}
  \caption{Architecture of the contextual TSR stage.}
  \label{fig:stage2}
  \vspace{-0.5cm}
\end{figure}

\subsection{Contextual target speaker refinement}

Although the coarse estimates preserve target identity, they may contain residual interference or missing target components. We therefore employ a TFGridNet-based TSR network that revisits the original mixture $y$. For each target, a contextual encoder jointly encodes $y$ and the corresponding coarse estimate $\tilde{s}_{ci}$; we adopt the contextual encoder from \cite{yang2024target}. The two contextual representations are concatenated and processed by stacked TFGrid blocks followed by a Deconv2D decoder:
\begin{equation}
    \left(\tilde{s}_{r1},\tilde{s}_{r2}\right)
    =F_{\mathrm{TSR}}\left(y,\tilde{s}_{c1},\tilde{s}_{c2}\right),
    \label{eq:stage2}
\end{equation}
where $\tilde{s}_{ri}$ is the refined estimate for target $i$. The mixture provides target components suppressed during coarse separation, while the two coarse estimates preserve target identity and provide complementary separation cues.

We supervise each refined estimate using its clean target $s_i$. The reconstruction loss combines SI-SDR with compressed complex-spectral reconstruction:
\begin{equation}
    \mathcal{L}_{\mathrm{rec}}
    =\sum_{i=1}^{2}\left[
    -\mathrm{SI\mbox{-}SDR}\left(\tilde{s}_{ri},s_i\right)
    +\lambda_r
    \left\|C_p\left(\tilde{S}_{ri}\right)-C_p\left(S_i\right)\right\|_2^2
    \right],
    \label{eq:recloss}
\end{equation}
where $\tilde{S}_{ri}$ and $S_i$ are the complex STFTs of the refined estimate and clean target, respectively.

To reduce over-suppression, let
$A_{i,tf}=\left|[C_p(S_i)]_{tf}\right|$,
$B_{i,tf}=\left|[C_p(\tilde{S}_{ri})]_{tf}\right|$, and let $W_{i,tf}$
indicate target-active time-frequency bins. We define the asymmetric loss
\begin{equation}
    \mathcal{L}_{\mathrm{over}}
    =
    \frac{
    \sum_{i,t,f}W_{i,tf}
    \left[\max\left(A_{i,tf}-B_{i,tf},0\right)\right]^2
    }{
    \sum_{i,t,f}W_{i,tf}+\epsilon
    },
    \label{eq:overloss}
\end{equation}
which penalizes missing target magnitude in active bins without directly encouraging excess energy.

We further employ a differentiable PESQ-oriented loss, implemented using \texttt{torch-pesq}\footnotemark, together with the adversarial training scheme of PASE~\cite{rong2026pase}.
\footnotetext{\url{https://github.com/audiolabs/torch-pesq}}
Let $\mathcal{D}=\{D_t,D_f\}$ contain the time- and frequency-domain
discriminators. The generator adversarial loss is
\begin{equation}
    \mathcal{L}_{\mathrm{adv}}
    =
    \sum_{D\in\mathcal{D}}\sum_{i=1}^{2}
    \mathbb{E}\left[
    \left(D(\tilde{s}_{ri})-1\right)^2
    \right],
    \label{eq:advloss}
\end{equation}
while the discriminators use the corresponding least-squares real/fake
objective. The overall refinement objective is
\begin{equation}
    \mathcal{L}_{\mathrm{TSR}}
    =
    \mathcal{L}_{\mathrm{rec}}
    +\mu_o\mathcal{L}_{\mathrm{over}}
    +\mu_p\mathcal{L}_{\mathrm{pesq}}
    +\mu_a\mathcal{L}_{\mathrm{adv}}
    +\mu_f\mathcal{L}_{\mathrm{fm}},
    \label{eq:stage2loss}
\end{equation}
where $\mathcal{L}_{\mathrm{fm}}$ denotes discriminator feature matching and
$\mu_o$, $\mu_p$, $\mu_a$, and $\mu_f$ are the corresponding loss weights.

\subsection{Post-fusion}

The coarse and refined estimates are complementary: the former preserves the reverberation and background ambience, whereas the latter favors cleaner target extraction. We linearly combine them as
\begin{equation}
    \tilde{s}_{fi}
    =(1-\alpha)\tilde{s}_{ci}
    +\alpha\tilde{s}_{ri},
    \qquad 0\leq\alpha\leq1,
    \label{eq:fusion}
\end{equation}
where $\tilde{s}_{fi}$ is the final estimate for target $i$ and $\alpha$ is a global fusion weight selected on the development set. The selected value is fixed for all test samples without reference-based or utterance-dependent tuning.

\begin{table*}[t]
\caption{Official test-set results in the overall (mix + remix) scope. Ours
denotes the submitted system; all competitors are anonymized. P808, SIG, and
BAK are display-only and do not enter the leaderboard OVRL. Higher is better
except CER and OVRL.}
\label{tab:test}
\centering
\setlength{\tabcolsep}{3.1pt}
\footnotesize
\resizebox{0.9\textwidth}{!}{%
\begin{tabular}{cc*{11}{c}}
\toprule
Track & System & SI-SDR & PESQ & STOI & UTMOS & P808 & SIG & BAK & DNS-OVRL & CER & SPK-SIM & OVRL\\
\midrule
\multirow{6}{*}{Track 1}
 & \textbf{Ours} & 10.70 & 2.93 & 0.84 & 2.07 & \textbf{2.93} & 2.34 & 2.62 & 1.83 & \textbf{0.123} & 0.75 & \textbf{2.86}\\
 & Team 1 & \textbf{12.72} & \textbf{3.00} & \textbf{0.85} & 2.10 & 2.71 & 2.14 & 2.42 & 1.70 & 0.145 & \textbf{0.77} & \textbf{2.86}\\
 & Team 2 & 9.16 & 2.77 & 0.83 & \textbf{2.15} & 2.81 & 2.36 & 2.48 & \textbf{1.94} & 0.135 & 0.74 & 3.57\\
 & Team 3 & 10.42 & 2.65 & 0.82 & 2.00 & 2.80 & 2.20 & 2.33 & 1.77 & 0.163 & 0.74 & 5.43\\
 & Team 4 & 10.03 & 2.63 & 0.82 & 2.00 & 2.78 & 2.19 & 2.25 & 1.73 & 0.177 & 0.73 & 6.29\\
 & Team 5 & 9.62 & 2.53 & 0.79 & 2.01 & 2.65 & 2.05 & 2.04 & 1.64 & 0.259 & 0.72 & 7.57\\
\midrule
\multirow{6}{*}{Track 2}
 & Team 1 & \textbf{12.26} & \textbf{2.94} & \textbf{0.84} & 2.13 & 2.68 & 2.11 & 2.39 & 1.67 & 0.153 & 0.76 & \textbf{2.43}\\
 & \textbf{Ours} & 10.22 & 2.89 & 0.83 & 2.09 & \textbf{2.88} & 2.26 & \textbf{2.49} & 1.76 & \textbf{0.148} & 0.74 & \textbf{2.86}\\
 & Team 2 & 7.05 & 2.56 & 0.78 & \textbf{2.17} & 2.77 & \textbf{2.30} & 2.42 & \textbf{1.90} & 0.220 & 0.71 & 5.00\\
 & Team 3 & 9.54 & 2.55 & 0.80 & 2.02 & 2.77 & 2.15 & 2.28 & 1.73 & 0.211 & 0.73 & 5.29\\
 & Team 4 & 9.27 & 2.59 & 0.80 & 1.87 & 2.70 & 1.84 & 1.91 & 1.52 & 0.206 & \textbf{0.76} & 5.29\\
 & Team 5 & 8.73 & 2.41 & 0.77 & 2.03 & 2.61 & 1.93 & 1.91 & 1.56 & 0.281 & 0.71 & 7.14\\
\bottomrule
\end{tabular}}
\end{table*}

\section{Experimental Setup}
\label{sec:experiments}

\subsection{Datasets}
During training, we use VoxCeleb2 \cite{chung2018voxceleb2} and the MISP 2023 Challenge dataset \cite{wu2024multimodal}. VoxCeleb2 provides diverse synchronized audio-visual utterances, while MISP introduces multi-party and far-field recording conditions. To construct each training mixture, we sample $M\in\{2,3,4\}$ speakers with probabilities 0.75, 0.20, and 0.05, respectively. We designate two speakers as visible targets and treat the remaining speakers as off-screen interferers. Each source utterance is independently convolved with a 16-kHz room impulse response sampled from OpenSLR SLR26 or SLR28 \cite{ko2017study}. We then sum the reverberant sources, add environmental noise, and crop the audio and corresponding video streams into synchronized 4.0-s segments.

We use official development sets for validation and parameter selection. The Track~1 development set contains 1,242 real-recorded \emph{mix} clips and 900 reference-available \emph{remix} clips, while the Track~2 development set contains 1,527 \emph{mix} clips and 1,098 \emph{remix} clips. We use these data for model selection and hyperparameter tuning, including the post-fusion coefficient.

We perform the final evaluation on the official test sets, which contain 4,257 clips for Track~1 and 4,941 clips for Track~2. The reference signals remain hidden and all test scores are computed by the organizers. Track~2 further introduces degraded visual conditions and additional far-field recordings, resulting in more severe visual corruption and acoustic mismatch than Track~1.

\subsection{Evaluation metrics}

We follow the official evaluation protocol of the ISCSLP 2026 Real-World AVSE Challenge~\cite{organizers2026iscslp2026realworldaudiovisual,yang2026identityfaithfulaudiovisualtargetspeaker}. SI-SDR~\cite{le2019sdr}, PESQ~\cite{rix2001perceptual}, and STOI~\cite{jensen2016algorithm} are evaluated on the \emph{remix} subset against clean references, whereas the non-intrusive UTMOSv2~\cite{saeki2022utmos,baba2024t05} and DNSMOS metrics are computed on both \emph{mix} and \emph{remix}. DNSMOS includes P.808 overall quality and P.835 SIG, BAK, and OVRL scores~\cite{reddy2022dnsmos,reddy2021dnsmos}. CER is computed from FunASR~\cite{gao2023funasr} hypotheses after text normalization, and SPK-SIM is the cosine similarity between WeSpeaker~\cite{wang2023wespeaker} embeddings of enhanced speech and enrollment voiceprints. Higher is better for all metrics except CER and the official OVRL rank score.

\subsection{Implementation details}

We train both the AVSS and TSR networks for 50 epochs using AdamW with an initial learning rate of $10^{-3}$, a weight decay of $10^{-4}$, and a batch size of 8. For progressive separation, we construct the intermediate targets using SIR levels of $\{0,5,10,15\}$ dB, followed by the interference-free coarse target. In the AVSS objective, we set the SI-SDR and compressed complex-spectral loss weights to $\lambda_s=\lambda_c=1.0$. For TSR, we assign unit weights to the SI-SDR and compressed complex-spectral reconstruction terms and set the over-suppression, PESQ-oriented, adversarial, and feature-matching loss weights to $\mu_o=0.1$, $\mu_p=0.05$, $\mu_a=0.01$, and $\mu_f=0.1$, respectively. For Track~2, we apply random visual degradation, including spatial masking, Gaussian blur, and frame dropout, and drop the visual modality with a probability of 0.1. We evaluate the post-fusion coefficient on the development set and select $\alpha=0.34$ according to the trade-off among intrusive quality, non-intrusive quality, recognition accuracy, and speaker similarity. The selected coefficient is fixed for both tracks during test inference.

\section{Results and Discussion}
\label{sec:results}

\subsection{Official test-set results}

Table~\ref{tab:test} compares our system with five anonymized teams on the official test set. Our system shares first place on Track~1 and ranks second on Track~2. It also achieves the lowest CER on both tracks, reaching 12.31\% on Track~1 and 14.79\% on Track~2. On Track~1, our method maintains competitive intrusive metrics together with strong speaker similarity. When evaluated under the more challenging visual conditions of Track~2, it exhibits only moderate degradation in SI-SDR, STOI, and SPK-SIM, suggesting that the coarse visual identity cues and mixture-conditioned TSR remain effective when the visual evidence becomes less reliable.

The results further demonstrate the multi-objective nature of real-world AVSE. Team~1 achieves better intrusive metrics, whereas several other systems obtain higher scores on individual non-intrusive quality estimators. In contrast, our method delivers consistently strong performance across speech quality, intelligibility, speaker preservation, and recognition accuracy. In particular, its leading CER results indicate that the proposed system effectively suppresses interference without severely damaging target-speech content.

\subsection{Development-set ablation study}

We conduct a compact ablation study on the Track 2 development set to evaluate the contributions of the TSR network and post-fusion. All variants are evaluated under the same protocol and differ only in the removed component.

\begin{table}[t]
\caption{Ablation results on the Track~2 development set.}
\label{tab:ablation}
\centering
\footnotesize
\setlength{\tabcolsep}{3.0pt}
\resizebox{\columnwidth}{!}{%
\begin{tabular}{lccccc}
\toprule
System & SI-SDR & PESQ & DNS-OVRL & CER & SPK-SIM \\
\midrule
Ours
& 11.53 & 2.65 & 1.86 & 0.117 & 0.68 \\
w/o post-fusion
& 9.29 & 2.17 & 2.20 & 0.132 & 0.57 \\
w/o TSR and post-fusion
& 10.88 & 2.55 & 1.56 & 0.129 & 0.66 \\
\bottomrule
\end{tabular}%
}
\vspace{-0.4cm}
\end{table}

Table~\ref{tab:ablation} demonstrate that coarse separation and TSR provide complementary estimates. TSR raises DNS-OVRL from 1.56 to 2.20, but decreases SI-SDR, PESQ, and SPK-SIM and increases CER, indicating that stronger perceptual enhancement may discard target-speech components. Post-fusion balances these behaviors, achieving the best SI-SDR, PESQ, CER, and SPK-SIM while preserving a clear DNS-OVRL improvement over coarse separation alone. The coarse and refined estimates act as correlated observations that retain different target information: the former preserves speaker identity and acoustic context, whereas the latter emphasizes perceptual quality. Combining them preserves their shared target content while recovering complementary components attenuated by either estimate alone.

\section{Conclusion}

We presented a two-stage method for real-world AVSE. Progressive, visually conditioned coarse separation establishes the two target identities while retaining their room responses and shared background noise. A mixture-conditioned TFGridNet-based TSR then produces cleaner target signals while mitigating over-suppression. We linearly fuse the coarse and refined outputs using a coefficient selected on the development set. The system ranked joint first on Track~1 and second on Track~2 of the ISCSLP 2026 Real-World AVSE Challenge and achieved the best CER on both tracks.


\bibliographystyle{IEEEtran}
\bibliography{mybib}

\end{document}